\documentclass[a4paper, amsfonts, amssymb, amsmath, reprint, showkeys, nofootinbib, twoside]{revtex4-1}
\usepackage[english]{babel}
\usepackage[utf8]{inputenc}
\usepackage{xcolor}
\usepackage[colorinlistoftodos, color=green!40, prependcaption]{todonotes}
\usepackage{wrapfig}

\usepackage{amsmath}
\usepackage{mathtools}
\usepackage{mathrsfs}

\usepackage{amsthm}
\usepackage{physics}
\usepackage{graphicx}
\usepackage[left=23mm,right=13mm,top=35mm,columnsep=15pt]{geometry} 
\usepackage{adjustbox}
\usepackage{placeins}
\usepackage[T1]{fontenc}
\usepackage{lipsum}
\usepackage{csquotes}

\usepackage{cancel}

\usepackage[normalem]{ulem}

\begin{document}

\title{Dynamically tuneable helicity in twisted electromagnetic resonators}
\author{E. C. I. Paterson} 
\email[]{22734222@student.uwa.edu.au}
\affiliation{Quantum Technologies and Dark Matter Labs, Department of Physics,  University of Western Australia, 35 Stirling Hwy, 6009 Crawley, Western Australia.}

\author{J. Bourhill}
\affiliation{Quantum Technologies and Dark Matter Labs, Department of Physics,  University of Western Australia, 35 Stirling Hwy, 6009 Crawley, Western Australia.}

\author{M. E. Tobar}
\affiliation{Quantum Technologies and Dark Matter Labs,  Department of Physics,  University of Western Australia, 35 Stirling Hwy, 6009 Crawley, Western Australia.}

\author{M. Goryachev}
\affiliation{Quantum Technologies and Dark Matter Labs,  Department of Physics,  University of Western Australia, 35 Stirling Hwy, 6009 Crawley, Western Australia.}

\date{\today} 

\begin{abstract}
    \noindent 
    We report the generation of helical electromagnetic radiation in a microwave cavity resonator, achieved by introducing mirror asymmetry, i.e., chirality, through a controlled geometric twist of the conducting boundary conditions. The emergence of this electromagnetic helicity is attributed to a nonzero spatial overlap between the electric and magnetic mode eigenvectors, quantified by $\text{Im}\left[\vec{\mathbf{E}}_i(\vec{r})\cdot{\vec{\mathbf{H}}}_i^*(\vec{r})\right]$, a feature not observed in conventional cavity resonators. This phenomenon originates from magnetoelectric coupling between nearly degenerate transverse electric (TE) and transverse magnetic (TM) modes, resulting in a measurable frequency shift of the resonant modes as a function of the twist angle, $\phi$. By dynamically varying $\phi$, we demonstrate real-time, macroscopic manipulation of both electromagnetic helicity and resonant frequency. In addition to the bulk helicity induced by global geometric twist, internal helical corrugations break structural symmetry on the surface, introducing an effective surface chirality $\kappa_{\text{eff}}$, which perturbs the resonant conditions and contributes to asymmetric frequency tuning. Furthermore, we investigate the underlying mode coupling dynamics of the system, highlighting strong photon-photon interactions.
\end{abstract}

\maketitle

\noindent 

\section{Introduction}

Chirality is a key feature in a broad range of physical systems, from particle physics \cite{Schw51,Adler69,Gooth:2017vg} to quantum and topological phenomena \cite{Ren:2022vd,Vu:2021wl,Gooth:2019np,Pikulin16,Wang13,Wieder:2022um,Fomin:2022aa} and complex molecular structures \cite{MacKenzie:2021vs,Tang2011,Torsi:2008wr,Hendry:2010ug,Cohen2010,Mun:2020ue}. In electromagnetism, chirality manifests through left-handed (LH) or right-handed (RH) polarisation states, which are intrinsically linked to the angular momentum of the electromagnetic field. The differential interaction of such radiation with chiral and non-chiral materials has garnered significant interest in fields such as materials science \cite{Zhaofeng_Chiral_Metamaterials}, nanophotonics \cite{Tang_Optical_Chirality,Zhao_Enhanced_Chiroptical}, and quantum information processing \cite{jonathan_Quantum,You_Quantum}, and the detection of dark matter or gravitational waves \cite{Bourhill_twisted_anyon_cavity_2023, Electromagnetic_Helicity,sym14102165}. 

The chirality of radiation can be described by electromagnetic helicity, a quantity intrinsically related to the dual symmetry of Maxwell's equations. Under a dual transformation, the electric and magnetic components of an electromagnetic field are mixed, leading to magnetoelectric coupling.

Mathematically, electromagnetic helicity, denoted $\mathscr{H}$, can be derived by projecting the complex electromagnetic state vector’s spin onto its linear momentum \cite{Alpeggiani_Electromagnetic_2018,PhysRevLett.113.033601,Martinez-Romeu:24,Bliokh_2013}. The sign of $\mathscr{H}$ indicates the handedness of the field. The expectation value of this operator provides a measure of the time averaged helicity density. 

For a monochromatic resonant mode $i$, the local, time-averaged helicity density may be written as
\begin{equation}
    h_i(\vec{r})=2\text{Im}\left[\vec{\mathbf{e}}_i(\vec{r})\cdot\vec{\mathbf{h}}_i^*(\vec{r})\right]=\frac{2\text{Im}\left[\vec{\mathbf{E}}_i(\vec{r})\cdot{\vec{\mathbf{H}}_i^*(\vec{r})}\right]}{V\mathcal{E}\mathcal{H}},
\label{eq:local_hel}
\end{equation}
where $\vec{\mathbf{E}}_i(\vec{r})=\mathcal{E}\,\vec{\mathbf{e}}_i(\vec{r})$ and $\vec{\mathbf{H}}_i(\vec{r})=\mathcal{H}\,\vec{\mathbf{h}}_i(\vec{r})$ are the electric and magnetic vector fields of the mode, respectively, $\mathcal{E}$ and $\mathcal{H}$ are real constants, and $V$ is the mode volume. The vectors $\vec{\mathbf{e}}_i(\vec{r})$ and $\vec{\mathbf{h}}_i(\vec{r})$ are the normalised position dependent eigenvectors (in a resonant system) such that $\frac{1}{V}\int \vec{\mathbf{e}}_i(\vec{r})^*\cdot\vec{\mathbf{e}}_i(\vec{r})dV=\frac{1}{V}\int\vec{\mathbf{h}}_i(\vec{r})^*\cdot\vec{\mathbf{h}}_i(\vec{r})dV=1.$

Integrating $h_i(\vec{r})$ over $V$ yields the total $\mathscr{H}$ of the mode,
\begin{equation}
    \mathscr{H}_i=\int h_i dV,
    \label{eqn:helicity_1}
\end{equation}
which provides a quantitative measure of the global chirality of the resonant electromagnetic field. While $\mathscr{H}_i$ characterises the overall handedness of a mode, $h_i(\vec{r})$ captures the local distribution of chirality within the cavity. It is easy to show that $\mathcal{E}=\sqrt{\frac{1}{V}\int |\vec{\mathbf{E}}_i(\vec{r})|^2 dV}$ and $\mathcal{H}=\sqrt{\frac{1}{V}\int |\vec{\mathbf{H}}_i(\vec{r})|^2 dV}$, making Eq.~\eqref{eq:local_hel} and Eq.~\eqref{eqn:helicity_1} consistent with other definitions of helicity~\cite{Bourhill_twisted_anyon_cavity_2023,Electromagnetic_Helicity}.

Electromagnetic helicity is typically observed in chiral surface states, at optical frequencies, or in complex meta-structures \cite{Liu_2014ur,Khanikaev_2013vy,Goryachev2016,Bliokh_Topological_2019}. It has been thought that an imaginary nonzero value of $\overrightarrow{\mathbf{E}}_i(\vec{r}) \cdot \overrightarrow{\mathbf{H}}_i^*(\vec{r})$ could not be generated in the free space of a cavity resonator~\cite{Tretyakov_waveguide_resonator_1995}. However, both here and in prior work~\cite{Electromagnetic_Helicity,Bourhill_twisted_anyon_cavity_2023}, resonant electromagnetic cavity modes with nonzero $\mathscr{H}$ in vacuo are generated by introducing mirror asymmetry into the cavity boundary conditions via a global geometric twist.

Beyond fundamental interest, helicity is being harnessed in practical technologies. Zheng et al. proposed a multi-channel single-pixel-imaging-based optical encryption scheme leveraging helicity as an additional degree of freedom to encode and multiplex meta-images on a single metasurface without requiring conventional digital key transmission~\cite{METASURFACE_ENCRYPTION}. This approach enhances encryption robustness, enables anti-counterfeiting, and introduces digital reconfigurability through distinct polarisation (helicity) states as encryption keys.

Huang et al.~\cite{HUANG2025177465} developed a strategy for improving radar stealth by controlling the helical phase of vortex waves using fixed helicity states in chiral phase-gradient metamaterials. Increasing helicity strengthens far-field diffusion and angular momentum characteristics of electromagnetic waves~\cite{Gauthier2017}, improving energy dispersion and attenuation~\cite{Li2021Programmable,Bai2023SmartMetasurface} and thereby reducing radar reflections~\cite{Zhang2018CodingMetasurface}. Such helicity control enables manipulation of scattering distributions, enhancing stealth performance and supporting secure communication.

We demonstrate that the frequency of helical modes can be continuously tuned by physically twisting the cavity resonators. Real-time modulation of $\mathscr{H}_i$ through mechanical deformation provides a capability that is difficult to achieve in photonic systems~\cite{Controlling_All_Optical_Helicity}. 

This dynamic tunability allows real-time control of microwave scattering and signal spreading, which may find uses in adaptive stealth responses and secure communication. In this context, dynamically tuneable $\mathscr{H}_i$ offers the possibility of helicity-modulated electromagnetic signals, which can serve as an additional physical-layer encoding channel~\cite{Zangeneh-Nejad2019}. By leveraging helicity-based keys that can be modulated on demand, interception and decoding may become more difficult, thereby enhancing adaptability and communication security.

Additionally, we demonstrate that internal RH helical corrugation along the internal walls of the device introduces an effective surface chirality, $\kappa_{\text{eff}}$, which generates RH electromagnetic helicity and induces measurable asymmetries in the resonant frequency response even in the absence of global twist. Beyond the generation of chiral electromagnetic modes, we show that geometric twisting also gives rise to strong coherent photon-photon coupling between these helical modes.

\section{Cavity Geometry}

\begin{figure}[t]
    \centering
    \includegraphics[width=1\columnwidth]{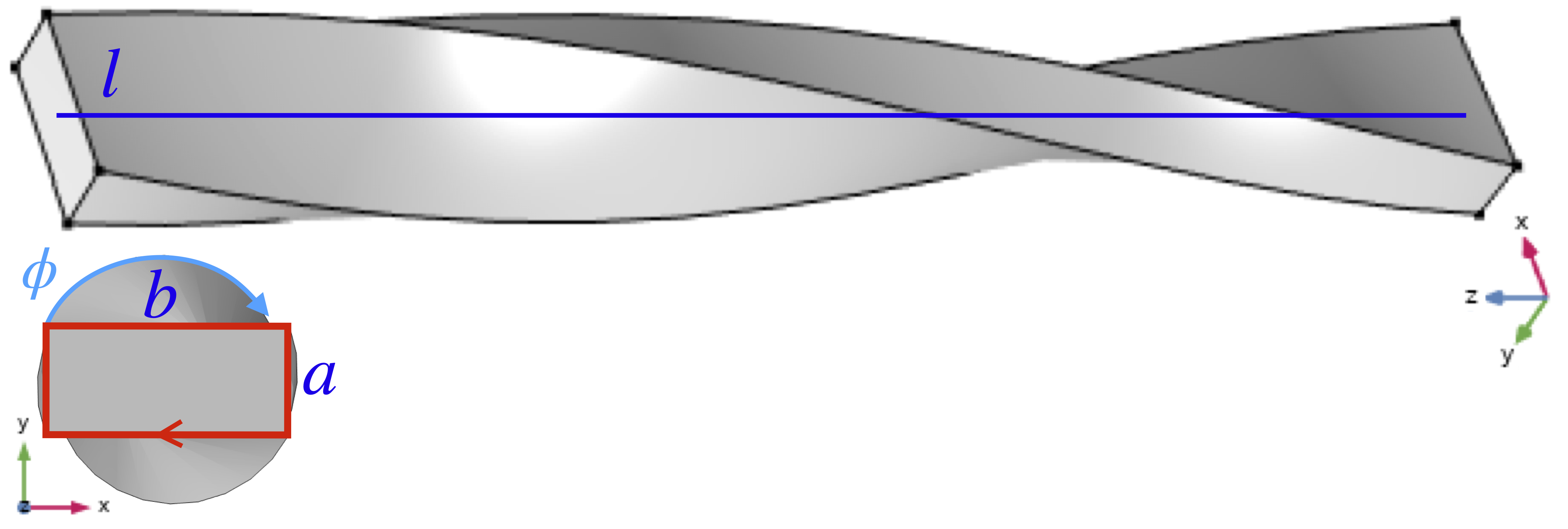}
    \caption{Geometry of the $\phi=\pi$ twisted cavity resonator with a WR-137 rectangular cross-section ($a=35.48$ mm, $b=16.1$ mm) and length $l=312.3$ mm.}
    \label{fig:geometry}
\end{figure}

We investigate a resonator with twisted electrically conducting boundary conditions. To construct this geometry, we take a rectangular prism and introduce a twist angle, $\phi$, perpendicular to the resonator’s central axis. The handedness of the twist is determined by the sign of $\phi$: a RH twist corresponds to $\phi > 0$, while a LH twist corresponds to $\phi < 0$. The resonator's geometry is represented in Fig.~\ref{fig:geometry}.

\section{Magneto-Electric Coupling and Mode Mixing}\label{sec:magneto_coupling}

The introduction of mirror asymmetry in the boundary conditions of the resonator results in a magneto-electric coupling that mixes a pair of near-degenerate transverse electric (TE) and transverse magnetic (TM) modes of the non-twisted cavity. This coupling transforms the orthogonal electric-magnetic field basis of the non-twisted resonator, $\mathbf{E}_0$ and $\mathbf{H}_0$, into a new basis $\mathbf{E}_i$ and $\mathbf{H}_i$ through a dual transformation~\cite{Alpeggiani18} (see Appendix~\ref{sec:mode_mixing_rect}). This mixing can be described by a transformation angle $\eta$, which governs the evolution of the field components~\cite{Alpeggiani_Electromagnetic_2018, Asker2018Axion, Planelles_Axion_2021}:
\begin{equation}
    \binom{\mathbf{E}_i}{c\mu_0 \mathbf{H}_i} =
    \begin{bmatrix}
        \cos(\eta) & \sin(\eta) \\
        -\sin(\eta) & \cos(\eta)
    \end{bmatrix}
    \binom{\mathbf{E}_0}{c\mu_0 \mathbf{H}_0}.
\end{equation}

This coupling lifts the degeneracy of the TE-TM pair, producing two hybridised eigenmodes, $\psi^+_{m,n,p}$ and $\psi^-_{m,n,p}$, which correspond to in-phase (TM$+$TE) and out-of-phase (TM$-$TE) superpositions of the original modes. These eigenstates may be written as
\begin{equation}
    \left|\psi^{\pm}_{m,n,p}\right\rangle = |\delta| \left|\text{TM}_{m,n,p} \right\rangle \pm |\beta| \left|\text{TE}_{m^\prime,n^\prime,p^\prime}\right\rangle,
    \label{eqn:state_equation}
\end{equation}
where the mode numbers $m$ and $n$ denote the number of transverse variations of the TM mode, and $p$ characterises its longitudinal structure. The primed indices $m^\prime$, $n^\prime$, $p^\prime$ describe the corresponding indices of the coupled TE mode. The coefficients $\delta$ and $\beta$ serve as weighting factors whose magnitudes evolve with increasing $\phi$, signifying progressive mixing between the TM and TE modes (see Appendix~\ref{section:weightings}). 

\begin{figure}[t]
    \centering
    \includegraphics[width=1\columnwidth]{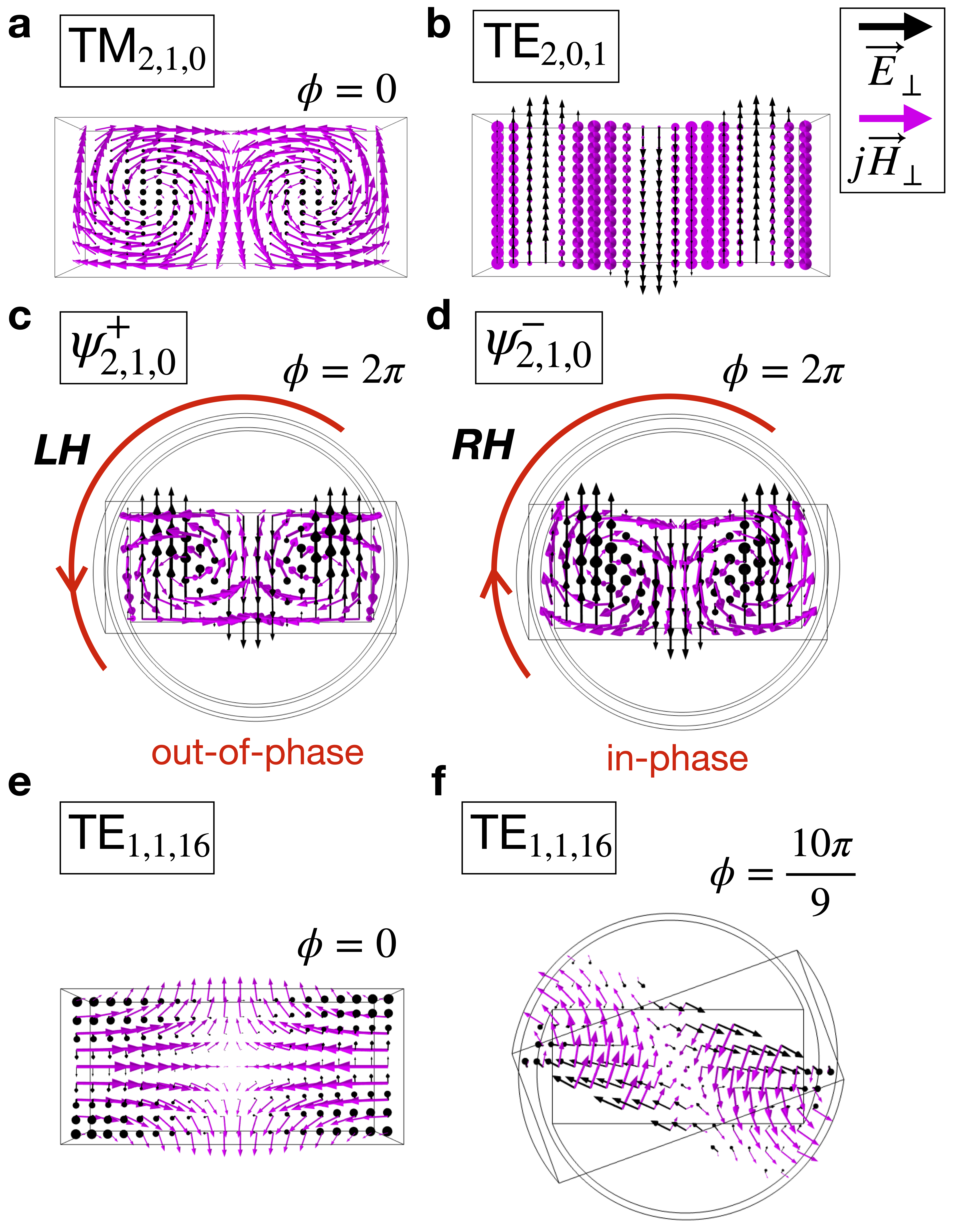}
    \caption{The $\vec{E}_\perp$ (magenta) and $j\vec{H}_\perp$ (black) field distributions for FEM eigenmode simulations of the WR-137 rectangular cross-section cavity resonator are shown for $\phi=0$ ((a) TM$_{2,1,0}$ and (b) TE$_{2,0,1}$ modes) and $\phi=2\pi$ ((c) $\psi_{2,1,0}^+$ and (d) $\psi_{2,1,0}^-$ modes). The corresponding fields for the TE$_{1,1,16}$ mode are shown for (e) $\phi = 0$ and (f) $\phi = \frac{10\pi}{9}$.}
    \label{fig:Modes}
\end{figure}

To demonstrate this mixing mechanism, we performed finite-element method (FEM) simulations of a resonator with a WR-137 rectangular cross-section of dimensions $35.48$ mm × $16.1$ mm and total length $l=312.3$ mm. These dimensions were chosen to match those of a commercially purchased corrugated twisted waveguide. The simulated transverse electric ($\vec{E}_\perp$) and magnetic ($\vec{H}_\perp$) field distributions, for the resonator with no net twist are shown in Fig.~\ref{fig:Modes}(a) and (b), corresponding to the $\mathrm{TM}_{2,1,0}$ and $\mathrm{TE}_{2,0,1}$ eigenmodes, respectively. When the resonator is twisted to $\phi=2\pi$, these modes couple. The out-of-phase mode mixing is shown in Fig.~\ref{fig:Modes}(c) generating the $\psi_{2,1,0}^+$ hybridised mode, and the in-phase mode mixing is shown in Fig.~\ref{fig:Modes}(d) generating the $\psi_{2,1,0}^-$ hybridised mode. The $\vec{H}_\perp$ fields now have an inherent handedness with respect to the $\vec{E}_\perp$ fields and hence we can classify the $\psi_{2,1,0}^+$ modes as LH and the $\psi_{2,1,0}^-$ modes as RH.

In our earlier work~\cite{Bourhill_twisted_anyon_cavity_2023}, it was emphasised that TE/TM degeneracy requires a cross-section belonging to the dihedral group of regular polygons, such as the $1{:}1$ square. This is indeed the case for the fundamental modes with $m=1$ and $n=1$, which are non-degenerate in a rectangular cross-section. However, for higher-order modes ($m\ge 2$ and/or $n\ge 2$), the standing-wave pattern divides the cross-section into $m\times n$ antinodal cells. In the rectangular resonator geometry shown in Fig.~\ref{fig:geometry}, the aspect ratio is close to $2{:}1$. In this case, the $m=2$ mode effectively partitions the cross-section into two nearly square subdomains ($a/2 \approx b$). Each subdomain effectively behaves like a local $1{:}1$ cavity. This sub-cell symmetry therefore produces TE/TM pairs with closely matched transverse wavenumbers, enabling the near-degeneracy required for strong mode mixing even in non-square geometries. We confirm this quantitatively using Eq.~\eqref{eqn:k_perp} obtaining $k_{\perp}=485.295~\mathrm{m}^{-1}$ for the TM$_{2,1,0}$ mode and $k_{\perp}=488.474~\mathrm{m}^{-1}$ for the TE$_{2,0,1}$ mode. These values correspond to a relative difference of only $\frac{\Delta k_\perp}{\bar{k}_\perp} \approx 0.655\%$.

\begin{figure}[t]
    \centering
    \includegraphics[width=1\columnwidth]{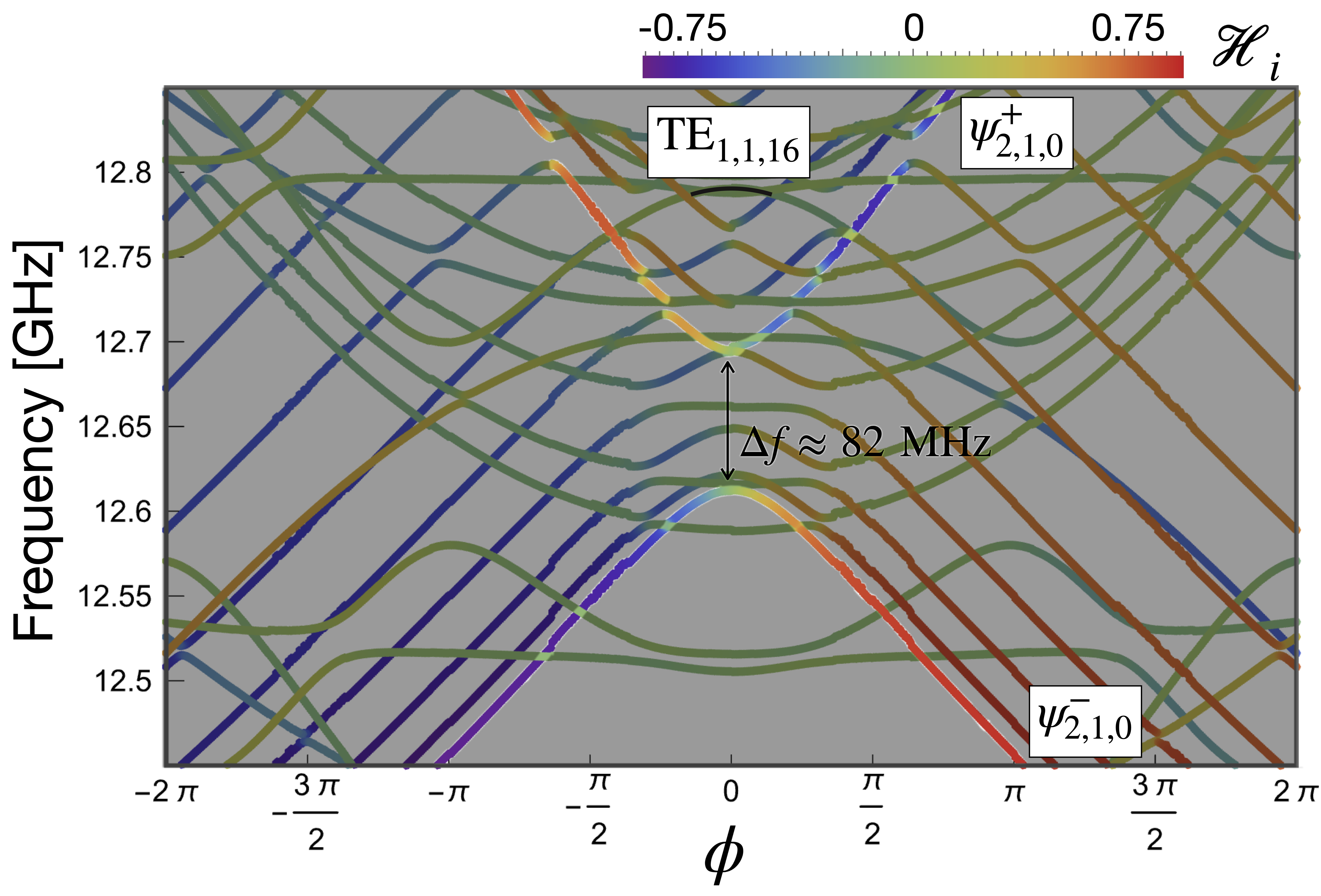}
    \caption{The eigenfrequencies $f_i$ of the resonant modes in a FEM simulated WR-137 rectangular resonator as a function of $\phi$. The solution colour represents $\mathscr{H}_i$. The frequency separation $\Delta f \approx 82~\text{MHz}$ between the TE$_{2,0,1}$ and TM$_{2,1,0}$ modes, which hybridise to form the helical states $\psi^\pm_{2,1,0}$ (illuminated), is marked by the arrow. The black line traces the TE$_{1,1,16}$ mode near $\phi = 0$.}
    \label{fig:eigenmodes}
\end{figure}

The generation of these helical modes was confirmed by parametrically twisting the FEM simulated WR-137 rectangular resonator shown in Fig.~\ref{fig:Modes}. The corresponding eigenfrequencies are plotted in Fig.~\ref{fig:eigenmodes} as a function of $\phi$, with color indicating $\mathscr{H}_i$. Notably, the $\psi^\pm_{m,n,p}$ eigenmodes rapidly acquire $\mathscr{H}_i$ even for small twist angles \(\phi\).

In addition to this TE-TM mode hybridisation mechanism, certain modes in the WR-137 cross-section resonator acquire nonzero $\mathscr{H}_i$ via a higher-order effect. Specifically, self-interference arising from the geometric asymmetry introduced by twisting the resonator can generate finite $\mathscr{H}_i$ in modes such as the TE$_{1,1,16}$ mode. The FEM-simulated $\vec{E}_\perp$ and $\vec{H}_\perp$ field distributions for this mode in the absence of twist are shown in Fig.~\ref{fig:Modes}(e). As the resonator is twisted to $\phi = \tfrac{10\pi}{9}$ (see Fig.~\ref{fig:Modes}(f)), the $\vec{E}_\perp$ and $\vec{H}_\perp$ components are no longer strictly orthogonal, indicating that the mode has acquired finite handedness.

This higher-order effect generates $\mathscr{H}_i$ at a much weaker rate than that produced via TE-TM mode hybridisation. The relative strength of these contributions can be quantified by comparing the sensitivity of $\mathscr{H}_i$ to small twist angles $\phi$, defined by $\left| \frac{d\mathscr{H}_i}{d\phi} \right|$ near $\phi = 0$. From the data in Fig.~\ref{fig:eigenmodes}, we find that near $\phi = 0$, the TE$_{1,1,16}$ mode exhibits $\left| \frac{d\mathscr{H}_i}{d\phi} \right| \approx 0.12~\mathrm{rad^{-1}}$, whereas the hybridised $\psi^{\pm}_{2,1,0}$ mode exhibits $\left| \frac{d\mathscr{H}_i}{d\phi} \right| \approx 1.08~\mathrm{rad^{-1}}$. Thus, the $\mathscr{H}_i$ generated via self-interference in the TE$_{1,1,16}$ mode is approximately an order of magnitude weaker than that generated by mode coupling in the $\psi^{\pm}_{2,1,0}$ modes.

\section{Helical Modes and Frequency Shifting}~\label{sec:hel_freq}

Not only do the $\psi_{2,1,0}^\pm$ modes in Fig.~\ref{fig:eigenmodes} acquire nonzero $\mathscr{H}_i$ under geometric twisting of the resonator, but they also exhibit symmetric frequency shifts relative to their non-twisted counterparts, TM$_{2,1,0}$ and TE$_{2,0,1}$ (see Appendix~\ref{section:rect_resonator_freq}). This frequency tuning closely resembles the behaviour observed in chiral materials.  

Material chirality is characterised by the dimensionless chirality parameter $\kappa$, which induces magnetoelectric coupling between the electric $\vec{\mathbf{E}}$ and magnetic $\vec{\mathbf{H}}$ fields according to:
\begin{equation}
\begin{aligned}
& \overrightarrow{\mathbf{D}} = \epsilon \overrightarrow{\mathbf{E}} - j \kappa \sqrt{\mu_0 \epsilon_0} \overrightarrow{\mathbf{H}}, \\
& \overrightarrow{\mathbf{B}} = \mu \overrightarrow{\mathbf{H}} + j \kappa \sqrt{\mu_0 \epsilon_0} \overrightarrow{\mathbf{E}},
\end{aligned}
\end{equation}
where $\kappa > 0$ and $\kappa < 0$ correspond to LH and RH polarisation rotation in the propagation direction, respectively. 

This symmetry-breaking term results in a nonzero $\mathscr{H}_i$, which is analogous to the mechanism responsible for generating nonzero $\mathscr{H}_i$ in a cavity resonator via geometric twisting (see Sec.~\ref{sec:magneto_coupling}). Accordingly, twisting the resonator is analogous to filling the cavity volume with an isotropic chiral medium characterised by the effective material parameter $\kappa \equiv \kappa_\text{eff}$, relative permittivity $\epsilon_r = 1$, and relative permeability $\mu_r = 1$, in that it reproduces the same $\mathscr{H}_i$. We emphasise that the equivalence between $\kappa_{\text{eff}}$ and $\phi$ is only true globally, and does not imply pointwise equivalence of $h_i$. Note that the persistence of the nonzero $\mathscr{H}_i$ in the resonator filled with a chiral medium arises from non-axial circulating field trajectories imposed by conducting boundary conditions (see Appendix~\ref{sec:fabre_perot}).

For a given resonator geometry and eigenmode, an empirical relationship between $\phi$ and $\kappa_{\mathrm{eff}}$ can be obtained by exploiting the fact that the $\mathscr{H}_i$ generated by geometric twisting can be nullified by an oppositely signed contribution induced by $\kappa_{\mathrm{eff}}$~\cite{Electromagnetic_Helicity}.

To identify this relationship, we extend the FEM simulations shown in Fig.~\ref{fig:eigenmodes} to a three-dimensional (3D) parameter space by twisting the WR-137 resonator while simultaneously sweeping the effective chirality parameter $\kappa_{\mathrm{eff}}$ introduced uniformly throughout the cavity volume. Eigenmodes for which $\mathscr{H}_i = 0$ correspond to points where $\phi$ and $\kappa_{\mathrm{eff}}$ are equal in magnitude and opposite in sign.

\begin{figure}[t]
    \centering
    \includegraphics[width=1\columnwidth]{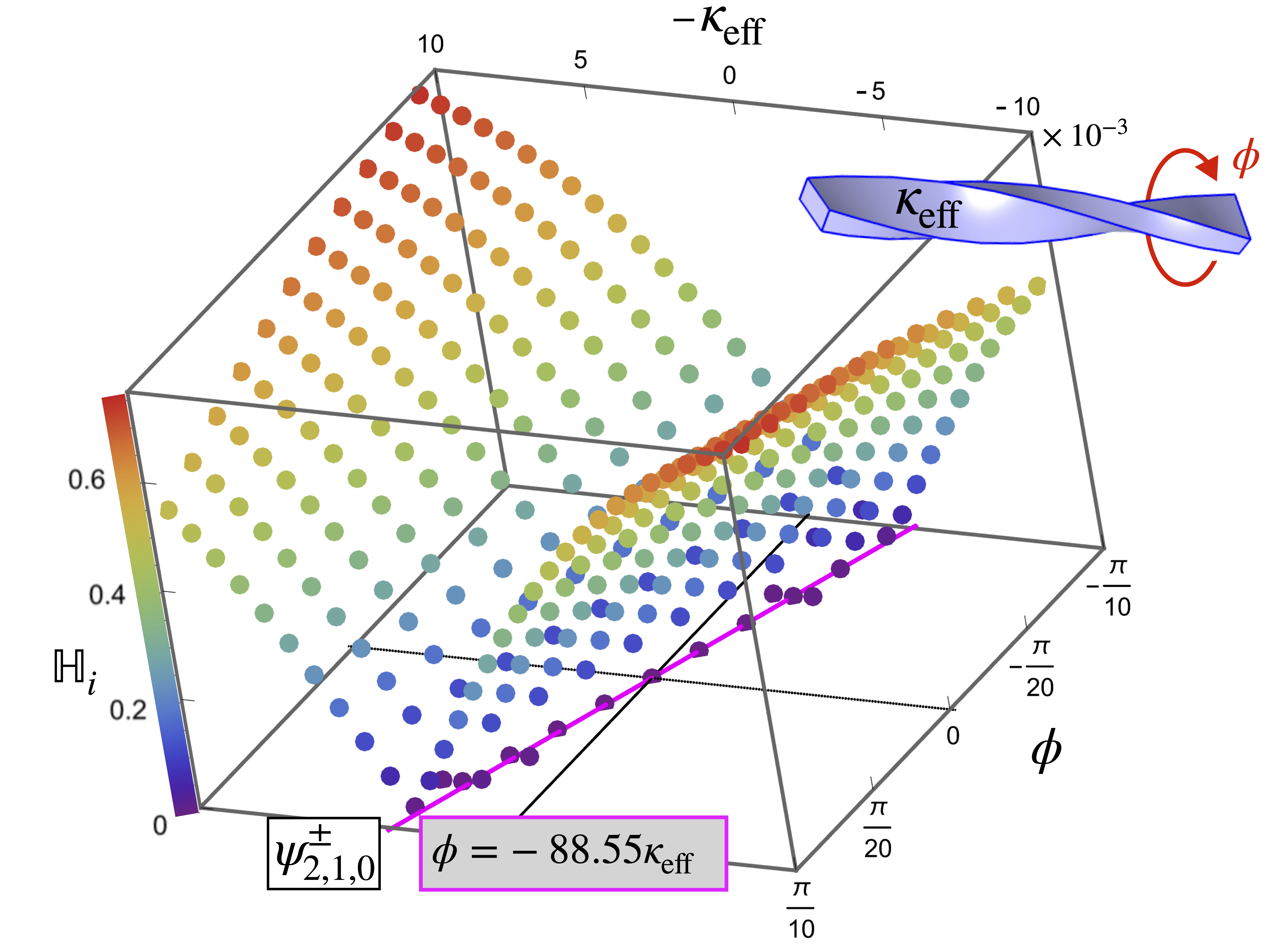}
    \caption{$\mathbb{H}_i$ derived from a FEM simulation of the $\psi_{2,1,0}^\pm$ modes in the twisted WR-137 resonator as a function of both 
    $\phi$ and $\kappa_\text{eff}$, 
    revealing null points where $\mathscr{H}_i$ induced by $\phi$ is cancelled by $\kappa_\text{eff}$.}
    \label{fig:kappa_equiv_plot}
\end{figure}

The eigenmodes obtained from these simulations for the helical mode $\psi^{\pm}_{2,1,0}$ are shown in Fig.~\ref{fig:kappa_equiv_plot} as a function of $\phi$ and $\kappa_{\mathrm{eff}}$. To clearly identify the de-hybridisation points, we plot the volume integral of the absolute helicity density,
\begin{equation}
    \mathbb{H}_{i}=\int |h_i| \, dV, \label{eq:abs_H}
\end{equation}
rather than $\mathscr{H}_i$, such that the null points appear as pronounced minima in the 3D parameter space where $\mathbb{H}_i = 0$.

By tracking these de-hybridisation points (indicated by the magenta curve in Fig.~\ref{fig:kappa_equiv_plot}) we extract the following effective, empirical relation for the $\psi^{\pm}_{2,1,0}$ mode:
\begin{equation}
    \kappa_\text{eff}^{(1)} \approx -\frac{\phi}{88.55}.
    \label{eqn:kappa_phi_psipm}
\end{equation} 
This relationship is mode-specific. For instance, for the TE$_{1,1,16}$ mode, the corresponding relationship is: 
\begin{equation}
    \kappa_\text{eff}^{(2)} \approx -\frac{\phi}{57.6}.
    \label{eqn:kappa_phi_0}
\end{equation}

\begin{figure}[t]
    \centering
    \includegraphics[width=1\columnwidth]{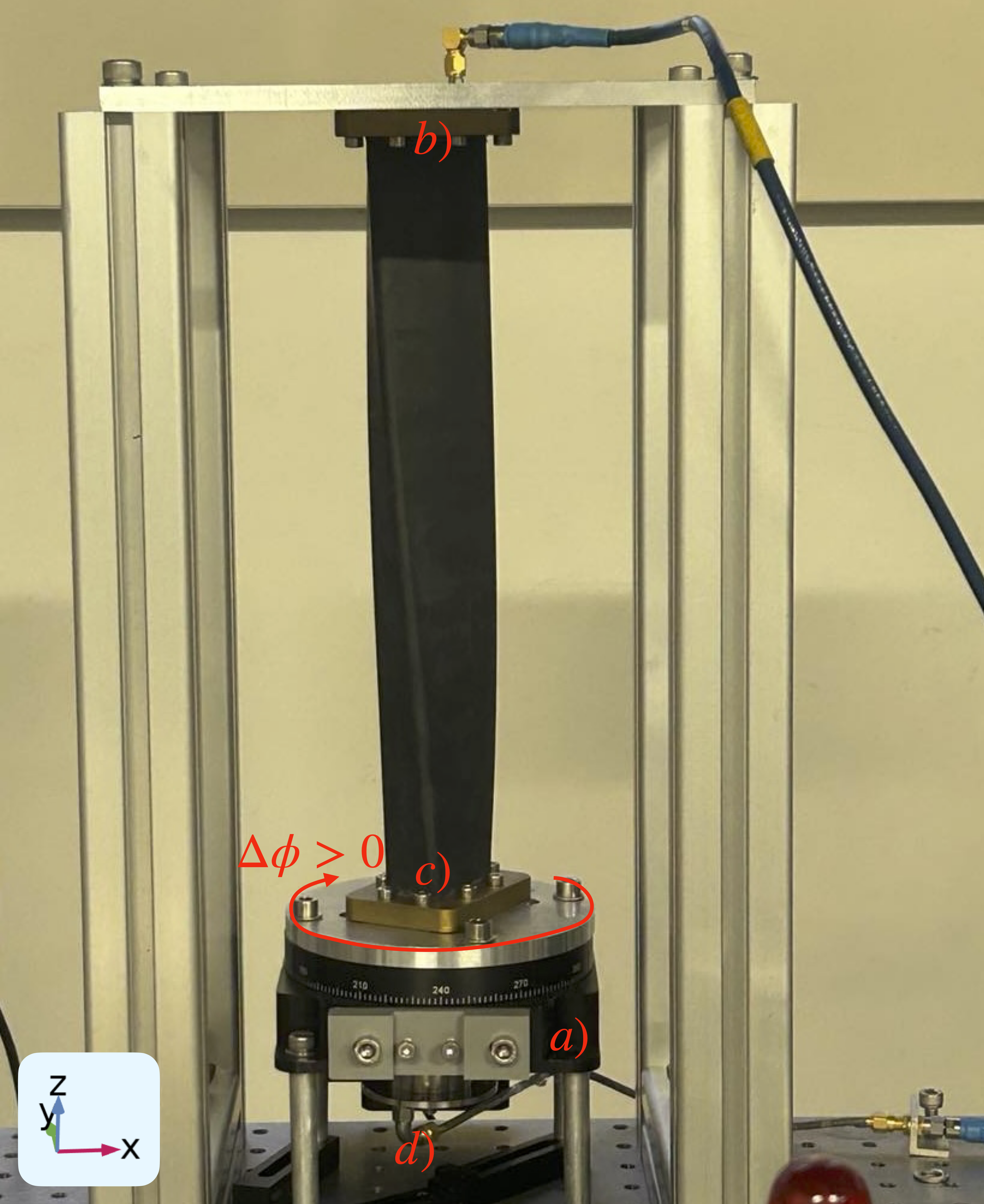}
    \caption{Experimental setup for inducing a controlled twist in a helically corrugated WR-137 rectangular resonator. (a) A rotary stage was used to apply mechanical rotation to one end of the resonator, while the other end remained fixed. The resonator was driven via coaxial probes inserted through the endcaps (b, c), with the lower probe (c) connected 
    through a rotating connector (d) to prevent cable torsion during rotation.}
    \label{fig:expr_set_up}
\end{figure} 

Having established an empirical relationship between $\phi$ and an equivalent $\kappa_\text{eff}$, we interpret the frequency shifts observed as $\phi$ is swept in Fig.~\ref{fig:eigenmodes} using the known relationship between $\mathscr{H}$ and the resonant frequency shifts resulting from variations in $\kappa$~\cite{Electromagnetic_Helicity}:
\begin{equation}
    \left(\frac{\delta \omega}{\delta \kappa \omega_0}\right)_{\kappa_0 \ll 1} = \frac{\mathscr{H}_0}{2 \mu_r \epsilon_r},
    \label{eqn:helicity_0}
\end{equation}
where $\mathscr{H}_0$ denotes the unperturbed $\mathscr{H}_i$, and $\delta \kappa = \kappa_1 - \kappa_0$ represents a small perturbation to the chirality parameter. Here, $\omega_1$ and $\omega_0$ are the perturbed and unperturbed resonant frequencies, respectively, and $\delta \omega = \omega_1 - \omega_0$. This proportionality highlights the sensitivity of the resonant frequency to $\mathscr{H}_i$ in bi-anisotropic systems. This relationship holds in both material systems, characterised by a chirality parameter $\kappa$, and in the non-material case of geometric twisting, because in both cases $\mathscr{H}_i$ is generated through magnetoelectric coupling.

\section{Dynamic Modulation of Helicity}
In 2014~\cite{Electromagnetic_Helicity}, it was demonstrated that frequency shifts arise due to mode mixing in triangular resonators twisted to discrete angles, thereby confirming the generation of $\mathscr{H}$ in a hollow, free-space volume as described by Eq.~\eqref{eqn:helicity_0}. These frequency shifts were observed to be symmetric about $\phi = 0$. Building on this result, we now demonstrate real-time frequency tunability by mechanically twisting a rectangular cross-section electromagnetic cavity resonator. 

\begin{figure}[t]
    \centering
    \includegraphics[width=0.9\columnwidth]{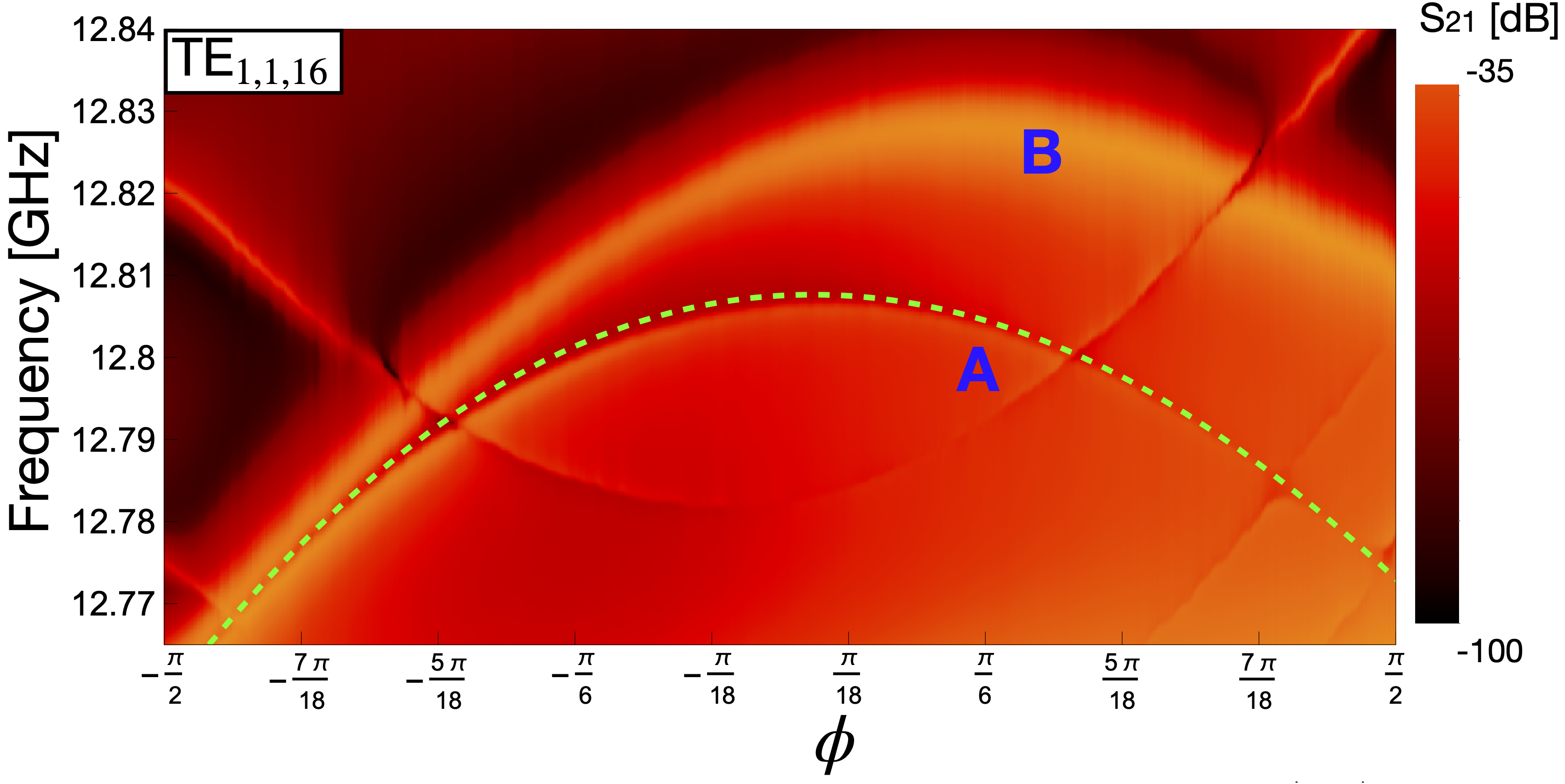}
    \caption{Transmission spectrum measured from the helically corrugated WR-137 rectangular cavity resonator. The yellow dashed line traces the frequency evolution of the hybridised mode, which arises from self-interference of the TE$_{1,1,16}$ mode under mechanical twist, and is labelled A. B labels a $\psi^\pm_{m,n,p}$ mode.}
    \label{fig:exper_freq_tuning_a}
\end{figure}

Although the rectangular configuration exhibits a lower $\mathscr{H}_i$ than the equilateral triangular case, due to coupling between modes with unequal in-plane propagation constants, which weakens hybridisation, it offers a practical advantage. Helically corrugated twistable waveguides of this type are commercially available, enabling straightforward experimental implementation of a continuously twistable resonator.

\begin{figure}[t]
    \centering
    \includegraphics[width=0.9\columnwidth]{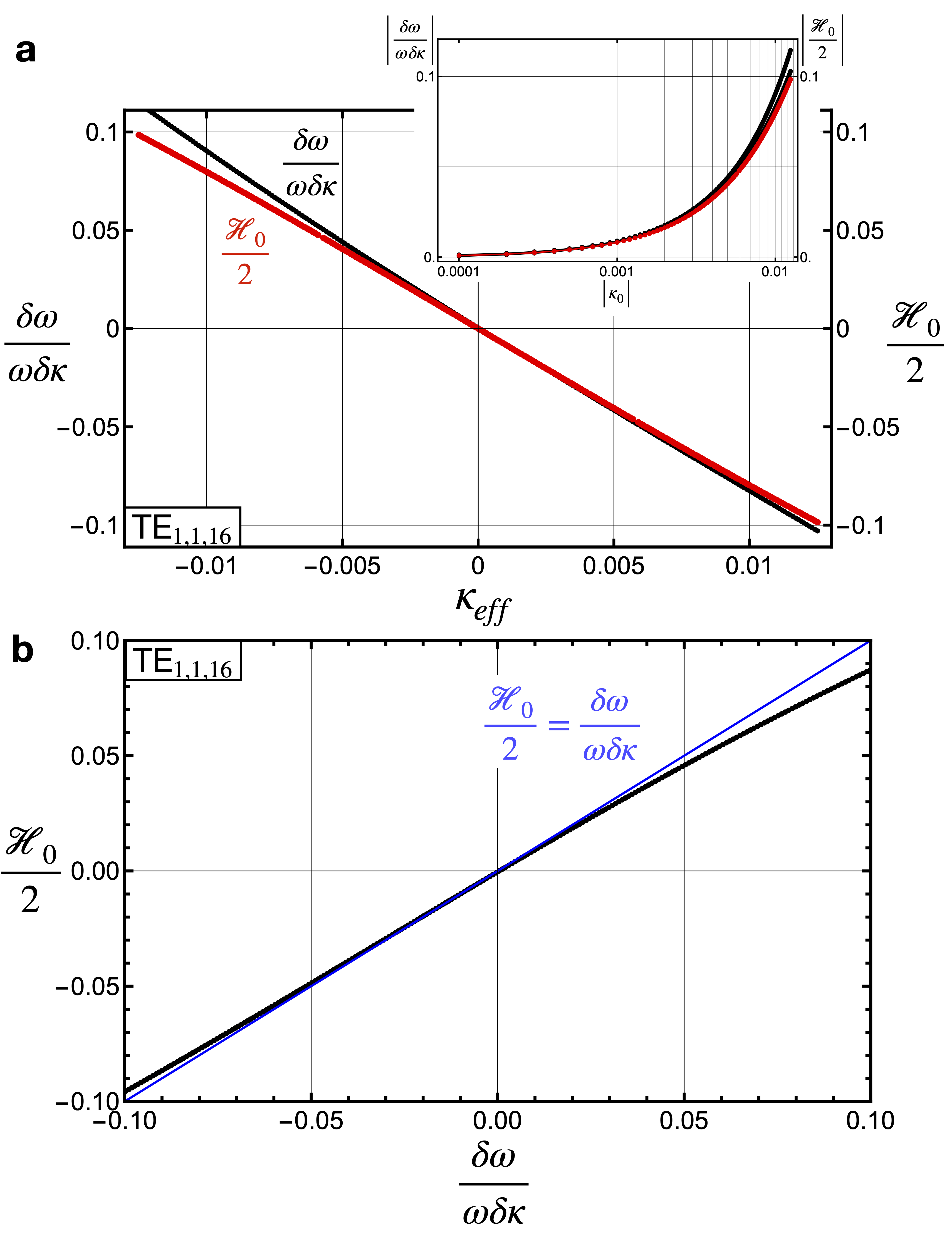}
    \caption{The hybridised mode frequencies traced in Fig.~\ref{fig:exper_freq_tuning_a} are used in (a) to plot the left-hand side of Eq.~\eqref{eqn:helicity_0} as a function of $\kappa_0$. FEM simulations of the non-twisted WR-137 rectangular resonator eigenmodes, incorporating an effective chirality parameter $\kappa_\text{eff}$ into the propagation medium, are used to calculate the right-hand side of Eq.~\eqref{eqn:helicity_0}, shown by the red line. The subplot presents the data on a log-linear scale. (b) Direct comparison between the simulated right-hand side and the experimentally derived left-hand side of Eq.~\eqref{eqn:helicity_0}. The theoretical relation, given by Eq.~\eqref{eqn:helicity_0}, is shown as a blue reference line.}
    \label{fig:exper_freq_tuning_2}
\end{figure}

We employed such a waveguide with dimensions matching those used in the FEM simulations informing Eqs.~\eqref{eqn:kappa_phi_psipm} and~\eqref{eqn:kappa_phi_0}, and formed the resonator by enclosing the open faces with metallic caps. One endcap was fixed, while the other was attached to a motorised rotary stage. The stage was synchronously controlled with a vector network analyzer (VNA), connected to the resonator via coaxial probes inserted through the endcaps. To prevent torsion in the VNA cables during rotation, a rotating connector was employed at the base end. Additionally, isolators were placed at both ports of the resonator to suppress line resonance effects during transmission ($S_{21}$) measurements. The resonator was quasi-statically twisted from a LH configuration ($\phi < 0$) to a RH configuration ($\phi > 0$), corresponding to a net positive twist angle $\Delta \phi > 0$. The experimental setup is provided in Fig.~\ref{fig:expr_set_up}. 

A portion of the measured transmission spectrum during mechanical twisting of the resonator is shown in Fig.~\ref{fig:exper_freq_tuning_a}, with the yellow dashed line tracing the TE$_{1,1,16}$ mode frequency as a function of twist angle. As expected, the twist-induced $\mathscr{H}_i$ leads to frequency tuning. It should be noted that the higher-frequency mode labelled B is a $\psi^\pm_{m,n,p}$ mode, distinguished by its larger frequency tuning with twist, consistent with the greater $\mathscr{H}_i$ expected for such modes. 

While the global frequency maximum is theoretically predicted at $\phi = 0$, it is experimentally observed at a positive twist angle of $\phi = 0.0832$ radians, indicating an asymmetry in the tuning response. This shift is attributed to internal RH helical corrugation, which breaks structural symmetry and induces an effective surface chirality. This chirality generates RH electromagnetic helicity, thereby perturbing the resonant conditions and producing measurable frequency shifts $\Delta\omega/\omega$, as described by Eq.~\eqref{eqn:helicity_0}. This effective surface chirality is inferred to be $\kappa_{\text{eff}} = -0.00144$ by applying Eq.~\eqref{eqn:kappa_phi_0} to the value of $\phi$ corresponding to the frequency maximum. The origin of this effect was verified using an independent resonant system incorporating RH helical corrugation, which exhibited a negative effective surface chirality consistent with the $\kappa_{\text{eff}}$ observed here, as detailed in Appendix~\ref{sec:corr_kappa_ver}. To account for this systematic asymmetry in the subsequent analysis of the TE$_{1,1,16}$ mode, a uniform offset of $-0.0832$ radians is applied to the transmission spectrum. 

According to the perturbative relationship given by Eq.~\eqref{eqn:helicity_0}, the observed frequency tuning induced by mechanical twisting implies tuning of $\mathscr{H}_0$. By employing the empirically derived relation between $\kappa_\text{eff}$ and $\phi$ (Eq.~\eqref{eqn:kappa_phi_0}), $\mathscr{H}_0$ can be computed for the TE$_{1,1,16}$ mode. This perturbation-derived $\mathscr{H}_0$ from experiment is plotted in black in Fig.~\ref{fig:exper_freq_tuning_2}(a). Also shown in red is $\mathscr{H}_0$ obtained from a FEM simulation of a non-twisted WR-137 resonator using Eq.~\eqref{eqn:helicity_1}, in which the effective chirality $\kappa_{\text{eff}}$ of an isotropic chiral material filling the resonator volume was swept. Good agreement is observed between simulation and experiment in the regime where $|\kappa_{\text{eff}}| \ll 1$, with deviations emerging for $|\kappa_{\text{eff}}| > 0.01$, as expected when the assumptions of perturbation theory begin to break down. 

\begin{figure}[t]
    \centering
    \includegraphics[width=1\columnwidth]{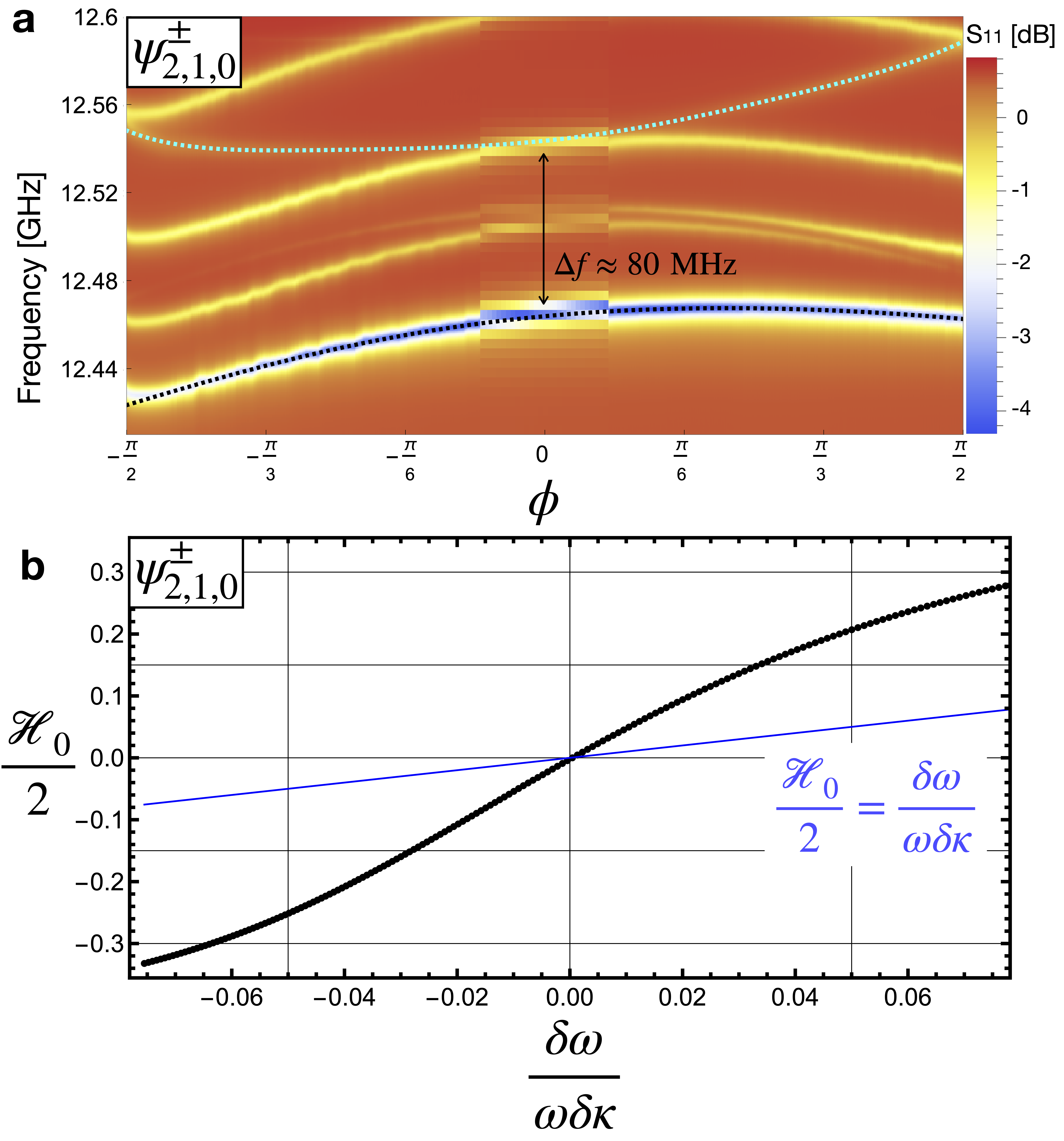}
    \caption{(a) Reflection ($S_{11}$) spectrum measured from the helically corrugated WR-137 rectangular cavity resonator. The frequency of the hybridised mode $\psi_{2,1,0}^\pm$, resulting from the mixing of the TE$_{2,0,1}$ and TM$_{2,1,0}$ modes, is indicated by the cyan and black dashed line. The experimentally measured frequencies of the $\psi_{2,1,0}^\pm$ mode traced in black is used in (b) to plot the left-hand side of Eq.~\eqref{eqn:helicity_0} against the right-hand side, which is obtained from FEM simulations of the non-twisted cavity eigenmodes incorporating an effective chirality parameter, $\kappa_\text{eff}$, into the propagation medium. The reference line corresponding to the theoretical relation Eq.~\eqref{eqn:helicity_0} is shown in blue.}
    \label{fig:exper_freq_tuning_psi}
\end{figure}

Note that the experimentally calculated $\mathscr{H}_0$ in the $\kappa_\text{eff}<0$ region are larger than that calculated in simulation due to corrugation induced asymmetry in frequency tuning. A direct comparison between the experimentally derived left-hand side and the simulated right-hand side of Eq.~\eqref{eqn:helicity_0} is shown in Fig.~\ref{fig:exper_freq_tuning_2}(b), revealing an approximately linear relationship between $\frac{\delta \omega}{\delta \kappa \omega_0}$ and $\mathscr{H}_0$, consistent with theoretical predictions. The small deviations from linearity reflect the previously noted corrugation effects. 

The TE modes generate $\mathscr{H}_i$ through a higher-order effect that is much weaker than the TE-TM mode hybridisation responsible for the generation of $\mathscr{H}_i$ in the $\psi_{m,n,p}$ modes, as discussed in Sec.~\ref{sec:magneto_coupling}. Consequently, the induced change in $\mathscr{H}_0$ of the TE$_{1,1,16}$ mode is small, resulting in a relatively modest frequency shift. In contrast, the $\psi_{2,1,0}^\pm$ mode generates a larger $\mathscr{H}_0$ under twisting, as will be shown experimentally. 

The reflection ($S_{11}$) spectrum for the $\psi_{2,1,0}^\pm$ modes is shown in Fig.~\ref{fig:exper_freq_tuning_psi}(a). The spectra recorded for $|\phi|\leq 0.24$ were taken with lower resolution in both frequency and twist. These modes can be identified as the $\psi_{2,1,0}^\pm$ states, arising from the coupling of the TM$_{2,1,0}$ and TE$_{2,0,1}$ modes, since at $\phi=0$ they exhibit nearly the same absolute frequencies as the corresponding eigenmodes in the FEM simulations of Fig.~\ref{fig:eigenmodes} (offset by $\approx 200$ MHz), and the frequency separation between them ($\Delta f \approx 80$ MHz) closely matches that found in simulation ($\Delta f \approx 82$ MHz). In the subsequent analysis we focus on the downwards-tuning $\psi_{2,1,0}^\pm$ modes, traced in black in Fig.~\ref{fig:exper_freq_tuning_psi}(a), namely $\psi_{2,1,0}^+$ for $\phi>0$ and $\psi_{2,1,0}^-$ for $\phi<0$. In the following analysis of these modes, the global frequency maximum has been shifted by $-0.682$ radians to align with $\phi = 0$, corresponding to a $\kappa_{\text{eff}}$ of $-0.00770$ via Eq.~\eqref{eqn:kappa_phi_psipm}. The sign of $\kappa_{\text{eff}}$ is consistent with that expected for RH helical corrugation.

Using the empirically derived relationship between $\kappa_\text{eff}$ and $\phi$ (see Eq.~\eqref{eqn:kappa_phi_psipm}), along with the experimentally measured mode frequencies, $\frac{\delta\omega}{\omega \delta\kappa}$ is calculated for the $\psi_{2,1,0}^\pm$ modes tracked in Fig.~\ref{fig:exper_freq_tuning_psi}(a). The right-hand side of Eq.~\eqref{eqn:helicity_0}, derived from simulation, is plotted against the perturbation-derived experimental values of $\frac{\delta\omega}{\omega \delta\kappa}$ in Fig.~\ref{fig:exper_freq_tuning_psi}(b). The figure reveals an approximately linear relationship between $\frac{\delta\omega}{\omega\delta\kappa}$ and $\mathscr{H}_0$, thereby supporting the validity of Eq.~\eqref{eqn:helicity_0}. While this relationship has a steeper gradient than the theoretical model (plotted in blue), this is due to discrepancies between simulation and experiment, such as the previously discussed effects of corrugation and experimental factors such as geometric changes arising from waveguide buckling during twisting. Despite these differences, the qualitative result remains: the sign of the rate of change of both $\mathscr{H}_0$ and $\frac{\delta\omega}{\delta\kappa \omega_0}$ is consistent, and their relationship remains approximately linear. We thus attribute the observed frequency shifts to the generation of $\mathscr{H}_i$.

\section{Strong Photon-Photon Coupling}\label{sec:strong_coupling}

\begin{figure}[t]
    \centering
    \includegraphics[width=1\columnwidth]{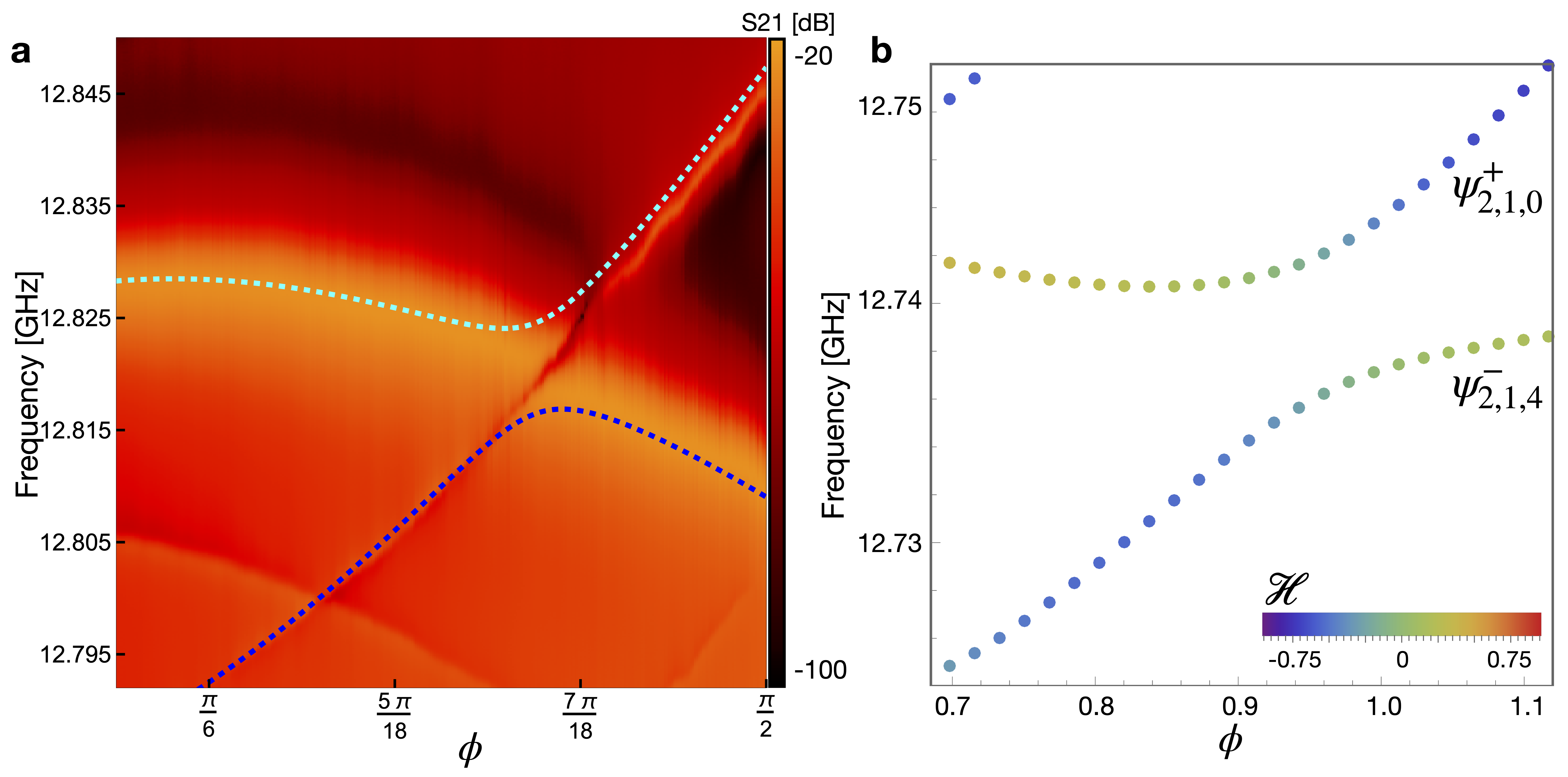}
    \caption{(a) Transmission spectrum of the helically corrugated WR-137 rectangular resonator as a function of $\phi$, centred on the mode crossing observed in the RH configuration. The dashed lines track the eigenfrequencies, $\omega_{\pm}$, resulting from the coupling between the $\psi_{2,1,0}^+$ and $\psi_{2,1,4}^-$ modes, obtained using Eq.~\eqref{eqn:mode_hybr_fit}. (b) FEM-simulated coupling of helical eigenmodes, resulting in cancellation of $\mathscr{H}_i$.}
    \label{fig:strongCouplingPlot}
\end{figure}

It can be seen from Fig.~\ref{fig:exper_freq_tuning_a} that two oppositely tuning helical modes, $\psi_{2,1,0}^+$ and $\psi_{2,1,4}^-$, intersect and exhibit an avoided level crossing, indicative of coupling. A zoomed-in view of the transmission spectrum around the hybridisation point of these helical microwave photon modes is shown in Fig.~\ref{fig:strongCouplingPlot}(a). The half-width at half-maximum dissipation rates of the un-coupled modes are $\frac{\Gamma_1}{2} = 0.588~\text{MHz}$ and $\frac{\Gamma_2}{2} = 1.950~\text{MHz}$. The interaction strength between the coupled modes is characterised by the coupling constant $g$, which is extracted from the minimum frequency splitting between them as a function of twist angle. In this system, the coupling constant is found to be $g = 4.05~\text{MHz}$.

The coupled eigenfrequencies, $\omega_{\pm}$, were calculated from the uncoupled frequencies of the $\psi_{2,1,0}^+$ and $\psi_{2,1,4}^-$ modes, $\omega_1$ and $\omega_2$, respectively, as a function of $\phi$ using the standard mode hybridisation model~\cite{PhysRevLett.121.137203,PhysRevApplied.22.064036}, given by
\begin{equation}
    \omega_{\pm} = \frac{\omega_1 + \omega_2}{2} \pm \frac{\sqrt{(\omega_1 - \omega_2)^2 + 4g^2}}{2}, 
    \label{eqn:mode_hybr_fit}
\end{equation}
and are overlaid on the transmission spectrum shown in Fig.~\ref{fig:strongCouplingPlot}(a). The cooperativity~\cite{PhysRevLett.113.156401}, defined as
\begin{equation}
    C = \frac{4g^2}{\Gamma_1 \Gamma_2},
\end{equation}
is calculated to be $C = 7.848 \gg 1$, indicating that the system operates in the strong-coupling regime. FEM simulations show that this type of hybridisation between helical modes leads to the cancellation of $\mathscr{H}_i$, as illustrated in Fig.~\ref{fig:strongCouplingPlot}(b). The hybridisation point appears at a different frequency in experiment than in simulation, owing to the corrugation-induced frequency asymmetry discussed earlier. This interaction point can therefore be utilised as an angle around which $\mathscr{H}_i$ can be rapidly modulated with minimal rotation.

\section{Conclusion}

Experimental results demonstrate that twisting an electromagnetic resonator facilitates real-time modulation of both electromagnetic helicity and resonant frequency. The introduction of a geometric twist induces magneto-electric coupling, which mixes electric and magnetic field components through a mixing angle $\eta$, resulting in the generation of chiral electromagnetic radiation in a monochromatic resonant mode \textit{in vacuo}. Moreover, internal helical corrugation introduces an effective surface chirality $\kappa_{\text{eff}}$, which leads to asymmetric frequency responses even in the absence of global twist, highlighting the role of structural details in shaping electromagnetic behaviour. The ability to dynamically modulate $\mathscr{H}_i$ enables a tuneable platform for helicity-modulated signals that can enhance communication security through providing an additional degree of freedom for physical-layer electromagnetic signal encryption. 

\section*{Acknowledgments}

\noindent This work was funded by the Centre of Excellence for Dark Matter Particle Physics, CE200100008. E.C.P is partially funded through the Defence Science Centre Research Higher Degree Student Grant.

\section*{Availability of Data}
\noindent All data and FEM model parameters are available upon request to the authors.

\providecommand{\noopsort}[1]{}\providecommand{\singleletter}[1]{#1}%

\newpage

\clearpage
\section*{Appendices} 
\addcontentsline{toc}{section}{Appendices} 

\setcounter{section}{0}
\setcounter{subsection}{0}
\setcounter{figure}{0}
\setcounter{table}{0}
\setcounter{equation}{0}

\renewcommand{\thesubsection}{A\arabic{subsection}}
\renewcommand{\thefigure}{A\arabic{figure}}
\renewcommand{\thetable}{A\arabic{table}}
\renewcommand{\theequation}{A\arabic{equation}}

\subsection{Mode Counting in the Hybrid Mode Basis}\label{sec:mode_mixing_rect}

\begin{figure}[t]
    \centering
    \includegraphics[width=1\columnwidth]{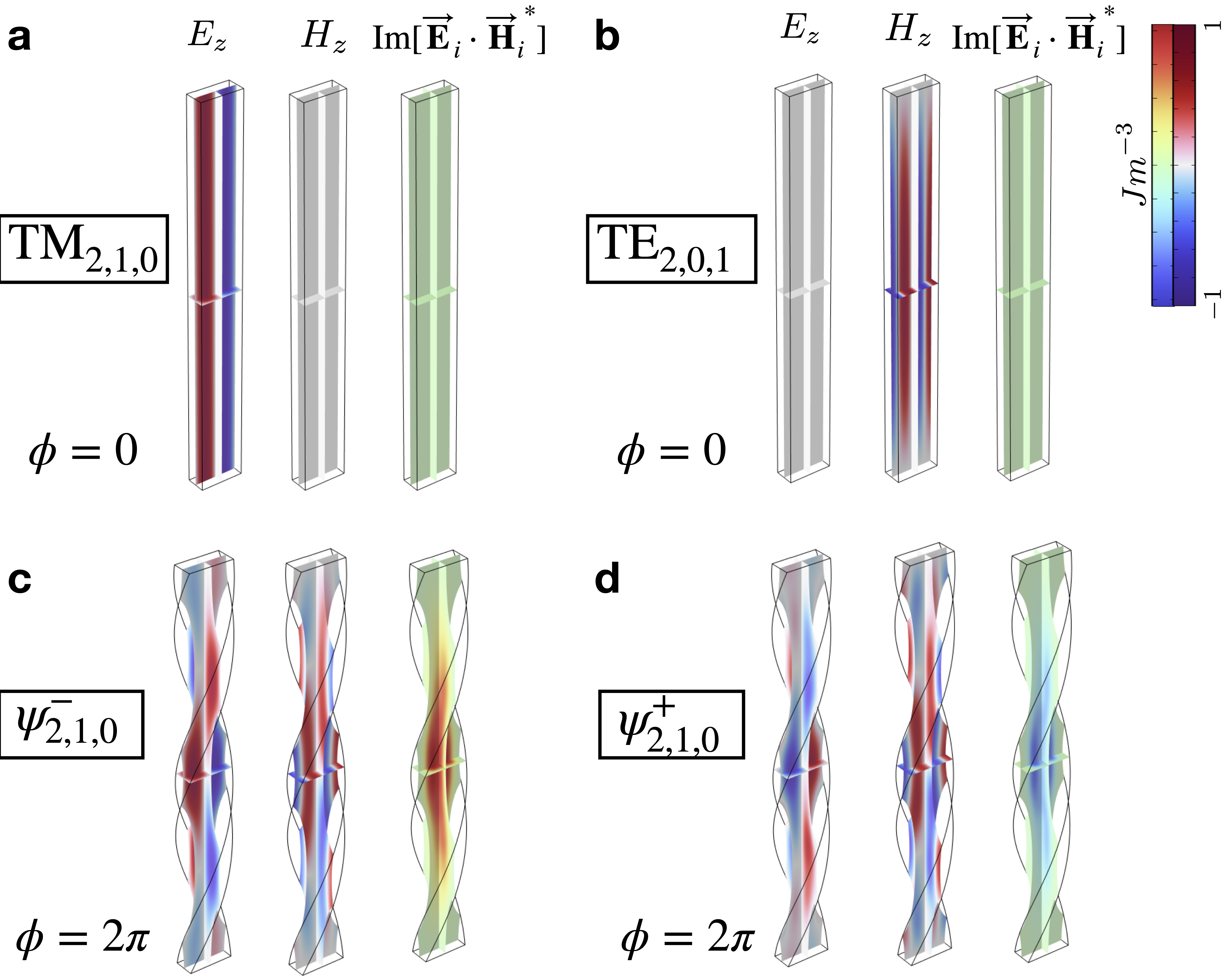}
    \caption{FEM-simulated normalised axial densities $E_z$, $H_z$, and $\text{Im}\left[\vec{\mathbf{E}}_i\cdot\vec{\mathbf{H}}_i^*\right]$, are shown for the (a) TM$_{2,1,0}$ and (b) TE$_{2,0,1}$ modes in the non-twisted WR-137 rectangular cross-section cavity resonator. Corresponding results for the hybrid (c) $\psi_{2,1,0}^-$ and (d) $\psi_{2,1,0}^+$ modes are shown for a twist angle of $\phi = 2\pi$. The longitudinal mode index $p$ can be identified by counting the number of maxima in the density plot of $|\text{Im}\left[\vec{\mathbf{E}}_i\cdot\vec{\mathbf{H}}_i^*\right]|$.}
    \label{fig:Rect_Modes}
\end{figure}

Figure~\ref{fig:Rect_Modes} depicts how the TE-TM mode mixing forms a new orthogonality basis. For the TM$_{2,1,0}$ and TE$_{2,0,1}$ modes the normalised field maxima can be counted by the normalised axial field densities $E_z$ and $H_z$ (see Fig.~\ref{fig:Rect_Modes}(a) and (b)), to determine the mode number $p$. However, this isn't the case for the hybrid $\psi_{2,1,p}^\pm$ modes (see Fig.~\ref{fig:Rect_Modes}(c) and (d)). As the $\vec{\mathbf{E}}_0$ and $\vec{\mathbf{H}}_0$ fields of the near-degenerate TE-TM modes mix due to the introduction of a twist the normalised field maxima can only be counted when evaluating the field product $\text{Im}\left[\vec{\mathbf{E}}_i\cdot\vec{\mathbf{H}}_i^*\right]$. The sign of this field product distinguishes the hybrid mode pair, being positive for $\psi_{2,1,p}^-$ and negative for $\psi_{2,1,p}^+$.

\subsection{Weighting Factors}\label{section:weightings}

The exact forms of the weighting factors $\delta$ and $\gamma$ used in Eq.~\eqref{eqn:state_equation} are 
\begin{equation}
    \delta = \frac{\int \mathbf{E}(\vec{r})_{\text{TM}_{m,n,p}} \cdot \mathbf{E}(\vec{r})_{\psi^\pm_{m,n,p}} d V}{\int \mathbf{E}(\vec{r})_{\text{TM}_{m,n,p}} \cdot \mathbf{E}(\vec{r})_{\text{TM}_{m,n,p}} \mathrm{dV}},
\end{equation}
and
\begin{equation}
    \beta = \frac{\int \mathbf{H}(\vec{r})_{\text{TE}_{m^\prime,n^\prime,p^\prime}} \cdot \mathbf{H}(\vec{r})_{\psi^\pm_{m,n,p}} d V}{\int \mathbf{H}(\vec{r})_{\text{TE}_{m^\prime,n^\prime,p^\prime}} \cdot \mathbf{H}(\vec{r})_{\text{TE}_{m^\prime,n^\prime,p^\prime}} \mathrm{dV}},
\end{equation}
where $\mathbf{E}(\vec{r})_{\psi^\pm_{m,n,p}}$ represents the electric field vector of the twisted $\psi^\pm_{m,n,p}$ modes, and similar notation applies for the magnetic field vectors. The same notation is used for the non-twisted modes.

\subsection{Frequency of the Non-twisted Rectangular Cavity Resonator}\label{section:rect_resonator_freq}

The frequencies of the resonant modes in the non-twisted resonators with a rectangular cross-section are given by:
\begin{equation}
f^R_{m, n, p}=\frac{c}{2} \sqrt{\left(\frac{m}{a}\right)^2+\left(\frac{n}{b}\right)^2+\left(\frac{p}{l}\right)^2},
\label{eqn:freq_untwisted_rect}
\end{equation}
where $c$ is the speed of light, $i \equiv m,n,p$ and $a$, $b$, and $l$ are the side lengths of the resonator, with $l > a > b$. The integers $m$ and $n$ denote the number of half-wavelength variations in the transverse dimensions, while $p$ represents the number of variations along the longitudinal axis. The selection rules for the TE modes are $m \geq n \geq 0$, $m \neq 0$, and $p > 0$, whereas for the TM modes, they are $m \geq n > 0$ and $p \geq 0$.

Equation~\eqref{eqn:freq_untwisted_rect} may equivalently be expressed in terms of the transverse wavenumber 
\begin{align} 
    k_\perp^2 &\equiv k_x^2 + k_y^2, \notag\\
    &= \left(\frac{m\pi}{a}\right)^2+\left(\frac{n\pi}{b}\right)^2, 
    \label{eqn:k_perp}
\end{align} 
and the longitudinal wavenumber 
\begin{equation}
    k_z \equiv \frac{p\pi}{l},
\end{equation} as \begin{equation} 
    f^{R}_{m,n,p}=\frac{c}{2\pi}\sqrt{k_\perp^2+k_z^2}.\label{eqn:freq_untwisted_k}
\end{equation}

\subsection{Persistence of Electromagnetic Helicity in Twisted Resonators with Effective Chirality}
\label{sec:fabre_perot}

\begin{figure}[t]
    \centering
    \includegraphics[width=0.9\columnwidth]{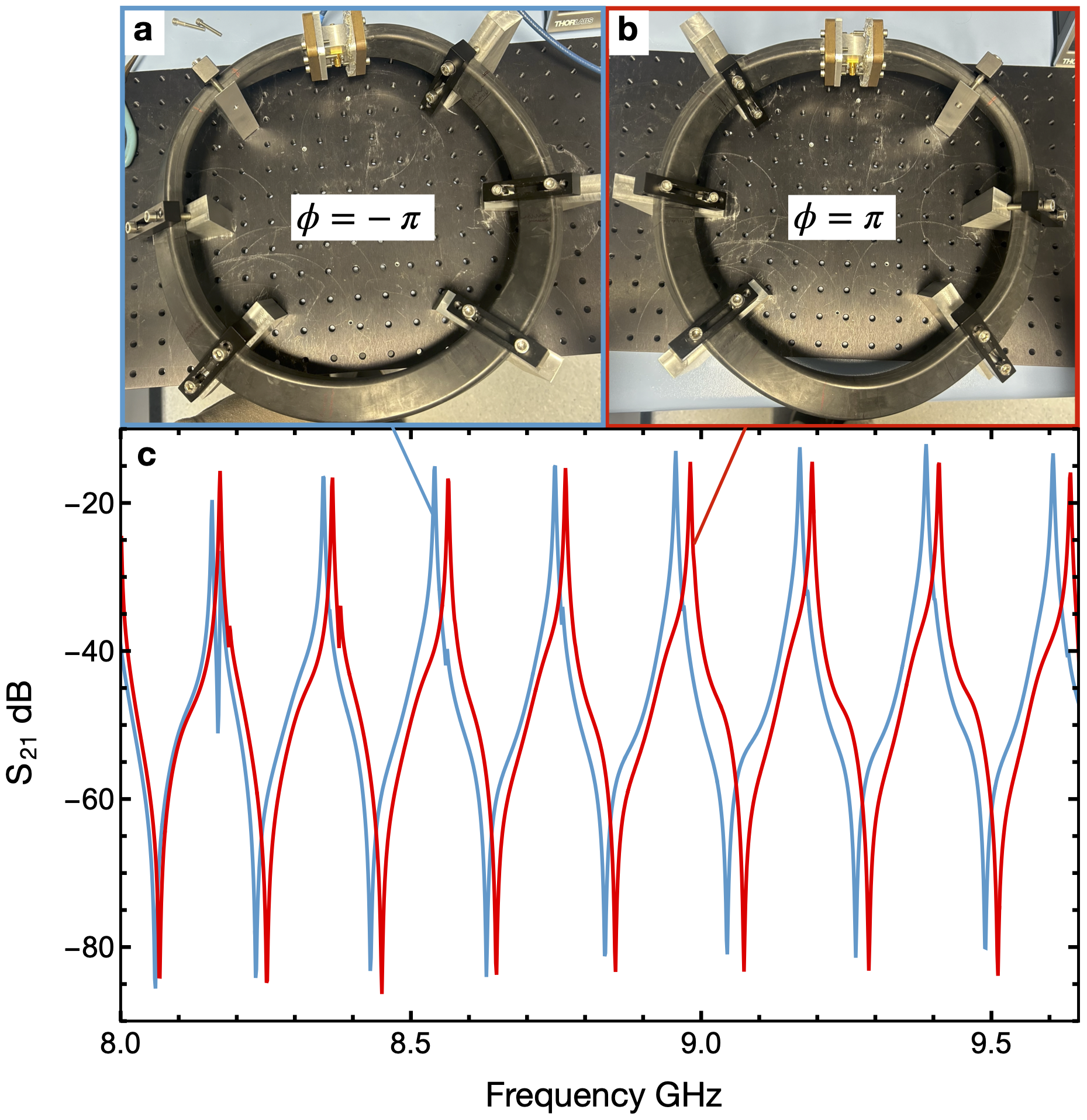}
    \caption{The experimental set-up of the (a) $\phi=-\pi$ and (b) $\phi=\pi$ M\"obius resonator cavities formed by bending a flexible RH helically corrugated WR-90 waveguide around on itself and clamping it to angled edges to ensure the correct twist angle. (c) Measured microwave transmission (S$_{21}$) spectrum of the TE$_{1,0,n}$ modes of the physically constructed resonators.}
    \label{fig:freq_shift_mobius}
\end{figure}

In conventional Fabry-Pérot cavities supporting purely axial modes, standing waves are formed from locally counterpropagating, time-reversed paths. Reflection at the cavity end mirrors reverses the axial wavevector and correspondingly flips the sign of $\mathscr{H}$. As a result, any $\mathscr{H}$ accumulated during a single pass through a chiral medium is exactly cancelled on the return pass, yielding zero net global $\mathscr{H}$, i.e. $\mathscr{H}_i = 0$.

The behaviour observed in twisted resonators differs fundamentally from this axial Fabry-Pérot picture. Owing to the conducting boundary conditions, the cavity eigenmodes are composed of globally circulating field trajectories that are not purely axial. In particular, the azimuthal component of the wavevector forms a closed loop that does not reverse upon reflection at the cavity end mirrors. Consequently, the $\mathscr{H}$ associated with this circulating component is preserved over successive round trips. As a result, when the twisted resonator is filled with an effective chiral medium characterised by $\kappa \equiv \kappa_{\mathrm{eff}}$, the cavity eigenmodes can sustain a net, nonzero global $\mathscr{H}$, i.e $\mathscr{H}_i\neq 0$.

\subsection{Verification of $\kappa_{\text{eff}}$ Induced by Helical Corrugation}
\label{sec:corr_kappa_ver}

To verify that the asymmetric frequency tuning observed under twisting for the resonant modes in the WR-137 rectangular resonators analysed in this work (hereafter referred to as the linear resonator system) arises from the RH helical corrugation of the device, rather than from other features, we examine the symmetry of the frequency tuning in an independent twisted system that also incorporates RH helical corrugation. Using numerical modelling, we further demonstrate that this asymmetric frequency tuning arises from the induction of an effective surface chirality $\kappa_{\text{eff}}$.

The system investigated consists of a flexible, helically corrugated WR-90 waveguide ($a = 22.86$ $\mathrm{mm}$, $b = 10.16$ $\mathrm{mm}$) configured as M\"obius resonators with opposite twist orientations, $\phi = -\pi$ and $\phi = \pi$. This M\"obius resonator system has been described in detail previously in Ref.~\cite{paterson2026dynamicallytuneablehelicitytwisted}. These configurations, illustrated in Fig.~\ref{fig:freq_shift_mobius}(a) and (b), were constructed by twisting the waveguide and closing the loop with a straight $85$-$\mathrm{mm}$-long section of waveguide housing two coaxial probes. Both M\"obius resonators had an approximate radius of $159$ $\mathrm{mm}$.

\begin{figure}[t]
    \centering
    \includegraphics[width=0.9\columnwidth]{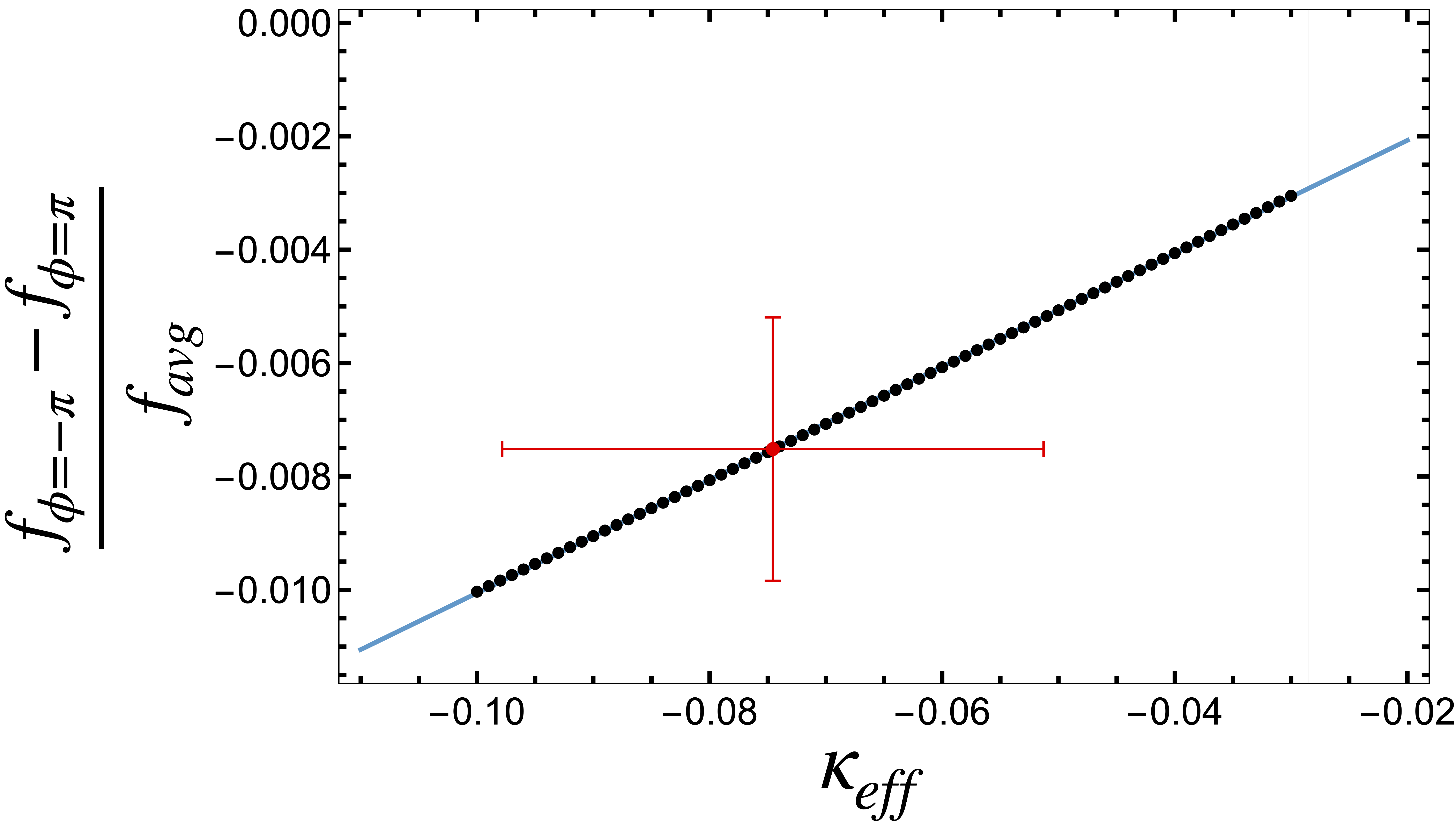}
    \caption{FEM-simulated relationship between the normalised frequency shift $(f_{\phi=-\pi} - f_{\phi=\pi}) / f_{\mathrm{avg}}$ and the effective chirality parameter $\kappa_{\text{eff}}$ for M\"obius resonators with RH helical corrugation. Black points represent simulated data with a linear fit shown in blue. The red point with error bars corresponds to the experimentally measured frequency shift between the physical $\phi = -\pi$ and $\phi = \pi$ corrugated M\"obius resonators shown in Fig.~\ref{fig:freq_shift_mobius}(c), yielding $\kappa_{\text{eff}} = -0.075 \pm 0.023$.}
    \label{fig:kappa_mobius_equiv}
\end{figure}

The measured transmission ($S_{21}$) spectrum of the TE$_{1,0,n}$ modes for the two twist configurations are shown in Fig.~\ref{fig:freq_shift_mobius}(c). In the absence of corrugation, the two configurations are expected to yield identical spectra, since twist-induced frequency tuning is ideally symmetric about $\phi = 0$. However, a clear frequency offset is observed. This demonstrates that both the corrugated M\"obius resonator system and the corrugated linear resonator system exhibit asymmetric frequency tuning under twisting. The observed asymmetry can therefore be attributed to their common feature, the helical corrugation, which breaks the symmetry about $\phi = 0$.

This frequency shift can be replicated in FEM simulations of the $\phi = -\pi$ and $\phi = \pi$ M\"obius resonators without corrugation by introducing an isotropic chiral material with effective chirality $\kappa_{\text{eff}}$ into the resonator volume. In fact, the normalised frequency shift between the two configurations, defined with respect to their geometric mean frequency $f_{\text{avg}}$, is linearly related to $\kappa_{\text{eff}}$ (see Fig.~\ref{fig:kappa_mobius_equiv}). Using this linear relationship and the experimentally measured normalised frequency shift between the $\phi = \pi$ and $\phi = -\pi$ corrugated M\"obius resonators, we extract a negative effective surface chirality associated with the RH helical corrugation:
\begin{equation}
    \kappa_{\text{eff}} = -0.075 \pm 0.023.
    \label{eqn:mobius_kappa}
\end{equation}
The experimental result is shown in red in Fig.~\ref{fig:kappa_mobius_equiv}. The experimental uncertainty in the normalised frequency shift is obtained from the standard error in the gradient of a least-squares linear fit to $(f_{\phi=-\pi} - f_{\phi=\pi})$ versus $f_{\mathrm{avg}}$, evaluated using a 95\% confidence interval. This gradient range is then mapped onto the simulation curve to determine the corresponding bounds on $\kappa_{\text{eff}}$. Therefore, although the magnitude of $\kappa_{\text{eff}}$ varies across different modes, the experimental uncertainty confirms that its sign is negative.

Introducing an effective surface chirality $\kappa_{\text{eff}}$ into the free space of the twisted resonant system without corrugation reproduces the asymmetric frequency tuning observed as the RH helically corrugated resonator system is twisted (see Figs.~\ref{fig:exper_freq_tuning_a} and \ref{fig:exper_freq_tuning_psi}(a)). This verifies that a corrugation-induced effective surface chirality is the underlying mechanism responsible for the observed asymmetry.

The value of $\kappa_{\text{eff}}$ (Eq.~\eqref{eqn:mobius_kappa}) for the RH helically corrugated M\"obius resonator differs in magnitude from that extracted for the linear RH corrugated resonator system. This difference is expected, as the two systems have distinct geometries, lengths, and aspect ratios, all of which influence the effective strength of the corrugation-induced chirality. Crucially, however, both RH helically corrugated systems induce a negative effective surface chirality $\kappa_{\text{eff}} < 0$, corresponding to the generation of RH electromagnetic helicity. This consistency in sign verifies that RH helical corrugation systematically induces a negative effective surface chirality, independent of the specific resonator geometry.

This analysis provides verification that the asymmetric frequency tuning observed in the twisted WR-137 resonators investigated in the main text can be attributed to the effective surface chirality $\kappa_{\text{eff}}$ introduced by RH helical corrugation. This mechanism explains the asymmetric frequency tuning of both the TE$_{1,1,16}$ and $\psi_{2,1,0}^\pm$ modes reported in this work.

\end{document}